# Critical assessment of hydrogen effects on the slip transmission across grain boundaries in α-Fe

I. Adlakha, and K.N. Solanki[*]

*School for Engineering of Matter, Transport, and Energy, Arizona State University, Tempe, AZ*

[*](480)965-1869; (480)727-9321 (fax), E-mail: kiran.solanki@asu.edu, (Corresponding author)

## 1. Abstract

Grain boundaries (GBs) play a fundamental role in the strengthening mechanism of crystalline structures by acting as an impediment to dislocation motion. However, the presence of an aggressive environment such as hydrogen increases the susceptibility to intergranular fracture. Further, there is a lack of systematic investigations exploring the role of hydrogen on the dislocation-grain-boundary (DGB) interactions. Thus, in this work, the effect of hydrogen on the interactions between a screw dislocation and <111> tilt GBs in α-Fe were examined. Our simulations reveal that the outcome of the DGB interaction strongly depends on the underlying GB dislocation network. Further, there exists a strong correlation between the GB energy and the energy barrier for slip transmission. In other words, GBs with lower interfacial energy demonstrate a higher barrier for slip transmission. The introduction of hydrogen along the GB causes the energy barrier for slip transmission to increase consistently for all of the GBs examined. The energy balance for a crack initiation in the presence of hydrogen was examined with the help of our observations and previous findings. It was found that the presence of hydrogen increases the strain energy stored within the GB which could lead to a transgranular-to-intergranular fracture mode transition.

*Keywords:* Grain boundaries; Slip transfer; Dislocation; Hydrogen embrittlement

## 2. Introduction

The mechanical properties and deformation behavior of many crystalline materials such as iron-based structural materials are controlled by the nature of grain boundaries (GBs), solute segregation, and subsequent dislocation-grain boundary (DGB) interactions. In particular, GBs play a fundamental role in determining the strengthening mechanisms as they act as an impediment to dislocation motion [1,2]. Additionally, the obstruction of dislocation glide can generate localization of strain that could eventually lead to a transgranular crack initiation [3], i.e., formation of slip bands that eventually lead to micro-shear crack nucleation. The presence of a corrosive environment such as hydrogen increases the tendency for a transition from a transgranular to an intergranular fracture mode [4–7]. For example, hydrogen has a strong bias to segregate around defects such as cracks, dislocations, and GBs [8,6,7,9,10]. The segregation of hydrogen along the GBs leads to a reduction in the cohesive strength [7,10], thereby promoting an intergranular failure [11–14]. However, there is a lack of systematic investigations exploring the role of aggressive environments (such as hydrogen, liquid gallium, and bismuth) and atomic-scale mechanisms of the DGB interactions [15]. For instance, the role of atomic structures along with hydrogen segregation on the pinning strength of the GB for the glissile lattice dislocations is a critical open question.

In the absence of a corrosive environment, there are four possible outcomes of the DGB interactions: a) a direct transmission; b) a direct transmission with a residual dislocation along the GB; c) an indirect transmission with a residual dislocation occurring because the incoming and outgoing slip planes do not intersect; and d) no transmission because the dislocation is absorbed at the GB. Several criteria [16–20] have been developed to predict the outcome for the DGB interaction. Shen et al. [18–20] devised a set of rules to predict the slip transmission when: a) the angle of intersection between both the incoming and outgoing slip plane is minimized and b) the resolved shear stress acting along the slip direction is maximized. Further, based on in-situ TEM experiments, Lee et al. [17] proposed an additional criterion to the model of Shen et al. [18–20] that ensures the outcome of slip transmission is based on the minimum residual dislocation along the GB, also known as the Lee-Robertson-Birnbaum (LRB) criteria. The geometric condition for the slip transmission can be expressed by the relative orientations of the slip planes and the GB plane normal in the following manner.

$$M = l_{in} \cdot l_{out} \qquad (1)$$

where, *l* is the unit normal at the intersection of the incoming and outgoing slip planes with the GB plane. Experimental verifications of these models are difficult because of microstructural heterogeneities introduced during processing which, along with dislocation pile-ups, can significantly alter the DGB interaction. Further, the atomic level details related to the interaction of individual GBs with dislocations still remain unclear due to insufficient resolution. For example, the GB atomic structures vary spatially which could change the pining strength of impeding dislocations. This has motivated several atomistic studies in BCC [21,22], FCC [23,24], and HCP [25] metals to validate these analytical models. In particular, previous atomistic [23,26] studies have shown that slip transmission was observed to violate the LRB criteria by transmitting on slip planes that did not have the highest resolved shear stress. Based on these findings, it was suggested that the underlying atomic structure at the DGB interaction site plays a key role in determining the outcome [23,24,26]. In spite of the large body of work on DGB interactions, there are still crucial open questions such as the effect of extreme environments on the slip transfer mechanism. Due to the high diffusivity, hydrogen atoms permeate through the microstructure rapidly by travelling along the GB networks and glissile dislocations (pipe diffusion) leading to mechanical degradation. In fact, the sensitivity of the GB to hydrogen embrittlement provides the impetus to explore the influence of hydrogen on the slip localization behavior.

In this paper, for the first time, we perform a systematic investigation of the role of hydrogen on the slip transmission mechanism in α-Fe using molecular dynamics. In particular, we study the effect of hydrogen on the interactions between the screw dislocation and eight <111> symmetric tilt grain boundaries (STGBs). Note that at low temperatures, the yield behavior of α-Fe is strongly governed by the glissile $a/2 < 111 >$ screw dislocation. Further, the observed outcomes for slip transmission were compared with the predictions of the LRB criterion. The next section in this paper describes the computational methods related to the GB and dislocation modeling and the method for the introduction of hydrogen along the GB. Then, we present several different characterizations of slip transmission with the implication of hydrogen embrittlement. For instance, our simulations of the DGB interactions reveal that the outcome depends strongly on the underlying GB dislocation network. Further, the energy barrier for slip transmission across the GB was quantified by measuring the strain energy flux across the DGB interaction site. A strong correlation (inverse power law relationship) was found between the static GB energy (measured at 0 K) and the energy barrier for the slip transmission with or without the hydrogen rich environment. In other words, the GB with lower interfacial energy demonstrated a higher barrier for the slip transmission, which is in agreement with previous experimental and modeling efforts. Thus, based on the present work and previous experimental studies [6,7,27,28], the implications of hydrogen embrittlement on intergranular fatigue crack initiation were examined. Hydrogen can affect plastic flow by lowering the energy barrier for dislocation slip, thereby enhancing strain localization [29,27,28], i.e., increasing the dislocation density within the pile-up. The increasing dislocation density within the pile-up also deposits additional hydrogen atoms at the GB causing a decrease in cohesive strength of the GB. Hence, a transition in failure mechanism, i.e., from transgranular-to-intergranular fracture mode, can occur due to energetically unfavorable conditions for slip transmission mechanisms. Therefore, the presence of hydrogen and its effects on the microstructural behavior can create circumstances for an intergranular crack initiation to occur at the pile-up site during fatigue loading conditions. However, the hydrogen enhanced localized plasticity (HELP) [27] is critical to the eventual transition to a intergranular fatigue crack initiation. Overall, these findings provide atomistic insights on the slip transferability and have general implications for the continuum scale crack initiation model development under extreme environments as exemplified by Tanaka and others [14,15,30,31].

## 3. Computational methods

Here, molecular dynamics calculations were performed using the large-scale atomic/molecular massively parallel simulator (LAMMPS) [32] platform to study the effect of hydrogen on the interaction of a screw dislocation with <111> STGBs (Figure 1) in α-Fe at 10 K. The semi-empirical embedded atom potential (EAM) developed by Ramasubramaniam et al. [33] was used to describe the Fe-H system, which is based on the Fe EAM potential of Hepburn and Ackland [34]. The Fe-H potential was parameterized using an extensive database of energies and configurations from density functional theory (DFT) calculations of various quantities including the dissolution and diffusion of hydrogen in bulk α-Fe, the binding of hydrogen to free surfaces, vacancies, dislocations, and other cross interactions between hydrogen and Fe. Moreover, the formation energies corresponding to multiple hydrogen-segregations to bulk α-Fe are consistent with the values predicted using ab initio calculations and experimentation [35]. The interatomic potential was able to accurately capture the threefold non-degenerate screw dislocation core in agreement with DFT findings [36,37], and the binding behavior of hydrogen around the dislocation core was found to be in agreement with DFT results [38]. Further, the interatomic potential has been widely utilized to study the effect of hydrogen on dislocation mobility [9,39], crack tip deformation [40], GBs [10,41,7], and surface energies [33,42] in Fe.

*3.1. Equilibrium grain boundary structures and energies*

The <111> STGBs selected for this work represent both the local minimum energy interfaces [43,44] and a large range of possible misorientation angles (refer to Figure 1 and Table 1). The GB structure and minimum energy were calculated using a bicrystal simulation cell with three-dimensional (3D) periodic boundary conditions consisting of two grains at 0 K as described by Rittner and Seidman [45]. The periodic boundaries (Y direction) were maintained with a separation distance of 12 nm between the boundaries. Several 0 K minimum energy GB structures were obtained through successive rigid body translations followed by an atom deletion technique and energy minimization using a non-linear conjugate gradient method [43–48]. The GB energies were found to be in good agreement with previous studies [10,43,52].

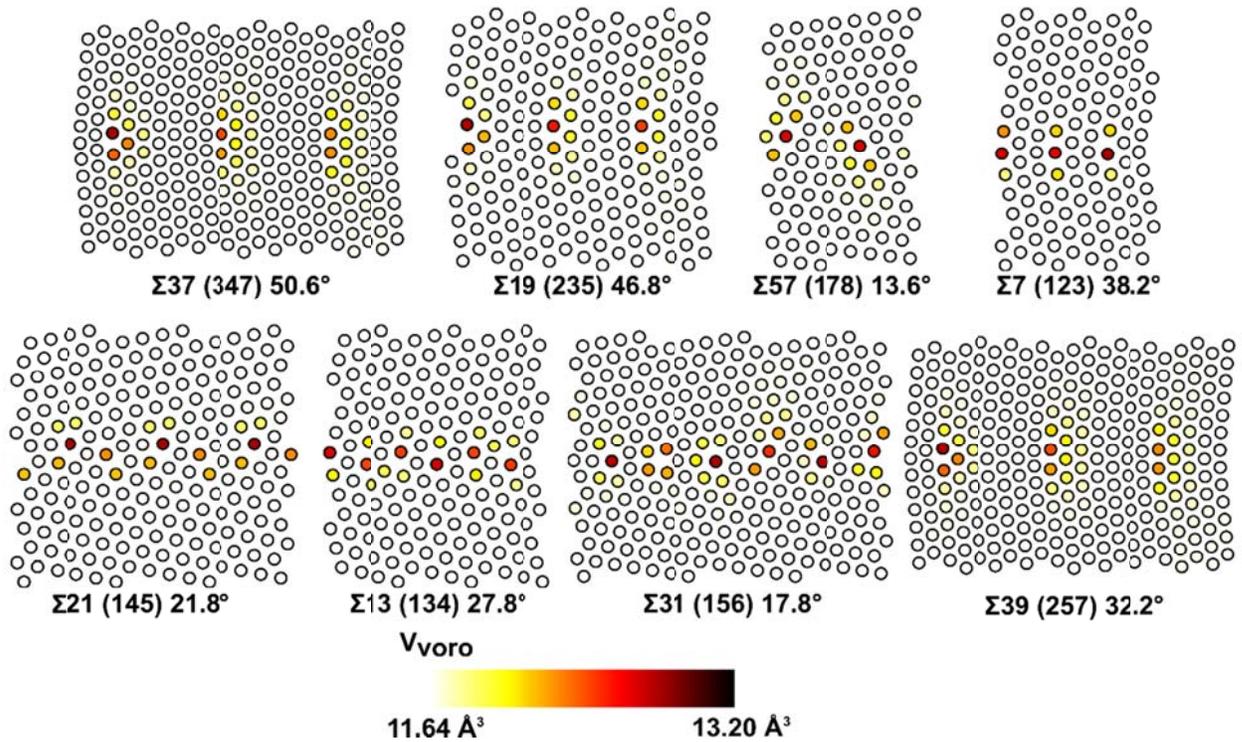

**Figure 1:** The <111> symmetric tilt GB structures in α-Fe used in this work. Atoms are colored based on the atomic volume estimated by Voronoi tessellation on a scale of 11.64 Å$^3$ to 13.20 Å$^3$.

**Table 1:** Grain boundary CSL description, misorientation angle, GB energy, the predicted and observed slip plane for the GBs examined in this work.

| CSL designation ($\Sigma$) | Misorientation angle, ($\theta$) | GB energy, (mJ/m$^2$) | Predicted outgoing slip plane | Observed outgoing slip plane |
|---|---|---|---|---|
| $\Sigma$37 (347) | 50.6 | 760 | $(1\bar{1}0)$ | $(1\bar{1}0)$ |
| $\Sigma$19 (235) | 46.8 | 887 | $(1\bar{1}0)$ | $(1\bar{1}0)$ |
| $\Sigma$57 (178) | 13.6 | 1030 | $(10\bar{1})$ | $(10\bar{1})$ |
| $\Sigma$7 (123) | 38.2 | 1056 | $(1\bar{1}0)$ | $(1\bar{1}0)$ |
| $\Sigma$21 (145) | 21.8 | 1114 | $(10\bar{1})$ | $(1\bar{1}0)$ |
| $\Sigma$13 (134) | 27.8 | 1117 | $(10\bar{1})$ | $(1\bar{1}0)$ |
| $\Sigma$31 (156) | 17.8 | 1127 | $(10\bar{1})$ | $(1\bar{1}0)$ |
| $\Sigma$39 (257) | 32.2 | 1137 | $(1\bar{1}0)$ | $(1\bar{1}0)$ |

*3.2. Simulation setup for the grain boundary-dislocation interaction in α-Fe*

In the present work, the bi-crystal simulation cell dimensions were approximately 300 Å × 300 Å × 40 Å. The orientation of the incident grain along the X [$\bar{1}\bar{1}2$], Y [$1\bar{1}0$], and Z [111] directions were fixed for all the DGB cases examined here (Figure 2a). The orientation of the transmitted grain was determined by the GB misorientation angle, refer to Table 1. The screw dislocation was introduced in the incident grain by applying Stroh's anisotropic displacement field [53] at a distance of approximately 50 Å away from the GB (see Figure 2a). The initial displacement field is depicted with the help of a differential displacement map in Figure 2b, which was found to be consistent with previous works [37,54]. Subsequently, the free boundary conditions were prescribed along the X and Y directions; while, a periodic boundary condition was maintained along the Z direction (Burger's vector direction). The atomistic model was then equilibrated at a temperature of 10 K using a canonical ensemble (NVT) for 5 ns, and the periodic direction along the Z direction was subsequently relaxed using the isothermal-isobaric (NPT) equations of motion for 15 ns. The top and bottom regions (~10 Å) along the Y direction were fixed (Figure 2a) and an incremental shear displacement was applied to the top of the atomic cell in the Z direction to obtain a constant shear strain rate ($\dot{\gamma}_{yz}$) of $10^8$ s$^{-1}$.

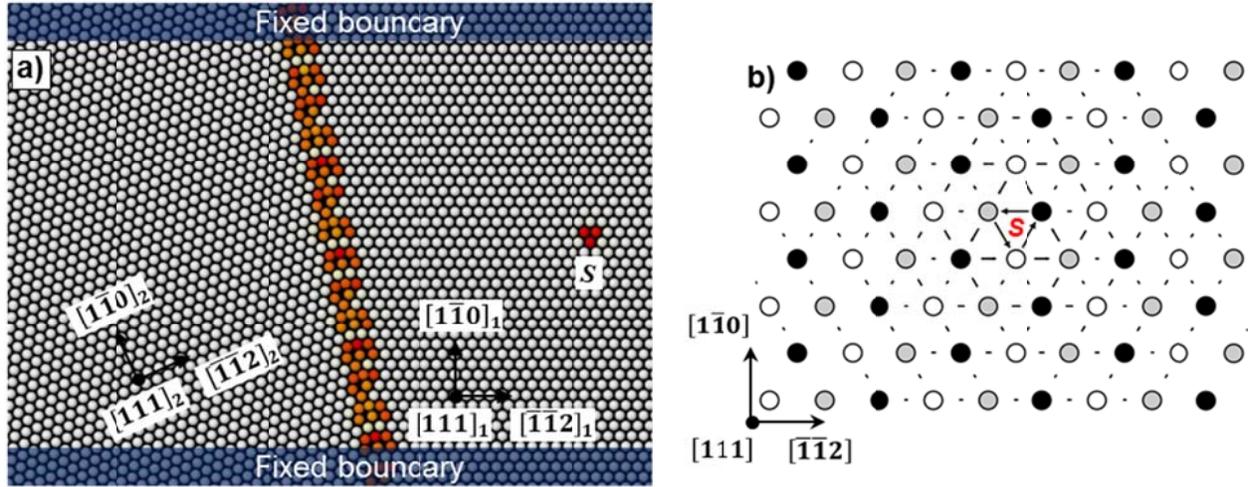

**Figure 2:** a) A schematic representation of the atomistic model employed to study the interaction between a screw dislocation (core marked as "S") and the <111> STGB in α-Fe. b) Differential displacement map of the compact core in Fe and the atomic colors emphasize different (111) planes.

*3.3. Introduction of hydrogen around the grain boundary*

The influence of hydrogen on the DGB interaction was examined by introducing hydrogen atoms around the GB. The hydrogen occupancy at a tetrahedral site ($\theta_i$) is dependent on the hydrogen binding energy ($E_b^i$) [55] and the temperature (*T*), which can be expressed in the following manner:

$$\frac{\theta_i}{1-\theta_i} = \frac{\theta_{bulk}}{1-\theta_{bulk}} exp\left(\frac{E_b^i}{k_B T}\right) \qquad (3)$$

where $\theta_{bulk}$ is the atomic fraction of hydrogen for the whole system, and $k_B$ is the Boltzmann constant. Note that the hydrogen atomic fraction of $3\times10^{-4}$ was chosen. The binding energy indicates the preference of a hydrogen atom to either remain at a particular GB site or a bulk lattice site. Negative values of the segregation energy implies that segregation is energetically favorable, while positive values indicate that it is favorable for hydrogen to leave the boundary and go into solution. A detailed survey on the site-to-site variation in the binding energy of hydrogen across several GBs can be found here [7,10]. Hydrogen atoms were initially placed along the GB based on the occupation probability for the GBs (Equation 3). Furthermore, the Monte-Carlo method was employed to obtain a realistic hydrogen distribution around the GB at a finite temperature (10 K).

*3.4. Quantifying the energy barrier for slip transmission*

The energy barrier for slip transmission across the GB was quantified by defining a control volume at the site of the DGB interaction [24]. The defected atoms (centrosymmetry parameter [56] > 0.5) within the control volume were used to estimate the energy barrier for slip transmission. The evolution of the net change in energy of the defected atoms during the loading process was quantified by comparing instantaneous energy ($E_t^i$) with reference/initial atomic energy ($E_r$). The net energy was normalized with the atomic volume occupied by defect atoms in the reference configuration:

$$E_{barrier} = \frac{\sum_i^n E_t^i - E_r}{V} \qquad (4)$$

where $V$ represents the total volume of the defected atoms within the control volume in the reference configuration.

**4. Results**

In this section, we discuss in detail the role of hydrogen on the atomic DGB interactions for a selected few GBs, such as Σ7 (123), Σ13 (134) and Σ57 (178). Note that eight GBs were used in this work, which represents both the local minimum energy interfaces [43] and a large range of possible misorientation angles. Further, isotropic elastic displacement fields of the screw dislocation and the <111> STGBs in α-Fe do not interfere with each other. However, due to the anisotropic nature of both of the defects at the atomic level, the dislocation may require thermal activation or critical stress to transmit across the interface. Therefore, we selected three boundaries to understand the precise role of the atomic details during slip transmission were visualized using various metrics such as the atomic shear strain invariant, common neighbor analysis, and centrosymmetry parameter in OVITO [57]. These three GBs (Σ7 (123), Σ13 (134) and Σ57 (178)) show a very distinct behavior in the presence of a hydrogen environment. For instance, in all three GBs with the addition of hydrogen, it was found that the underlying structure of the GB was distorted; therefore, the energy barrier for slip transmission increases. Furthermore, the Σ13 (134) STGB shows change in the transmission mechanism, i.e., from a direct to indirect transmission, while the Σ7 (123) STGB shows no noticeable change in the transmission mechanism (i.e., indirect transmission with or without aggressive environments).

*4.1. Σ7 (123) STGB: No noticeable change in the transmission mechanism*

The sensitivity of hydrogen embrittlement to GB structure provides the impetus to explore how variations in atomic environment in the vicinity of a GB influence the slip transmission. Towards this, we first examine the atomic events during a screw dislocation interaction with the Σ7 (123) GB. In the absence of a hydrogen environment, as the applied shear strain increases, the screw dislocation overcomes the Peierls stress and begins to glide towards the GB (Figure 3a). At an applied shear strain ($\gamma_{yz}$) of 3%, the screw dislocation was absorbed into the GB (Figure 3b) at a site of high atomic volume along the GB, see Figure 3b inset which is a magnified view of the atomic Voronoi volume variation due to the formation of the GB (Figure 1). The anisotropic elastic effects of the screw dislocation can be easily accommodated at this site by rearrangement of the GB dislocation network. Subsequently, the dislocation was transmitted across the GB at an applied shear strain ($\gamma_{yz}$) of 5.7% (Figure 3c). The dislocation was transmitted across the GB into the neighboring grain through an indirect manner, in other words, the GB sites for the dislocation absorption and transmission were separated from each other by 6 Å along the GB (Figure 3c). The dislocation was transmited on the $(1\bar{1}0)_2$ plane, which is in agreement with the outgoing plane predicted by the LRB criteria (Figure 3d). The energy barrier for the slip transmission was $2.75 \times 10^{11}$ mJ/m$^3$.

The addition of hydrogen atoms along the GB distorts the underlying GB structure, thereby decreasing the coincident sites along the GB. This behavior can be quantified by the shear strain required to transmit the dislocation which was much higher in the presence of hydrogen ($\gamma_{yz}$ = 8.8%) (Figure 3e). The magnified view in Figure 3d shows hydrogen atom distributions around the dislocation-GB interaction site. The energy barrier for the dislocation transmission increases from $2.75 \times 10^{11}$ mJ/m$^3$ for a hydrogen free case to $4.16 \times 10^{11}$ mJ/m$^3$ in the presence of hydrogen, i.e., a 51% increase in the energy barrier. However, one intersesting observation is that there is no noticeable change in the transmission mechanism, i.e., indirect transmission with or without hydrogen and at the same separation distance of 6 Å between the dislocation absorption and transmission.

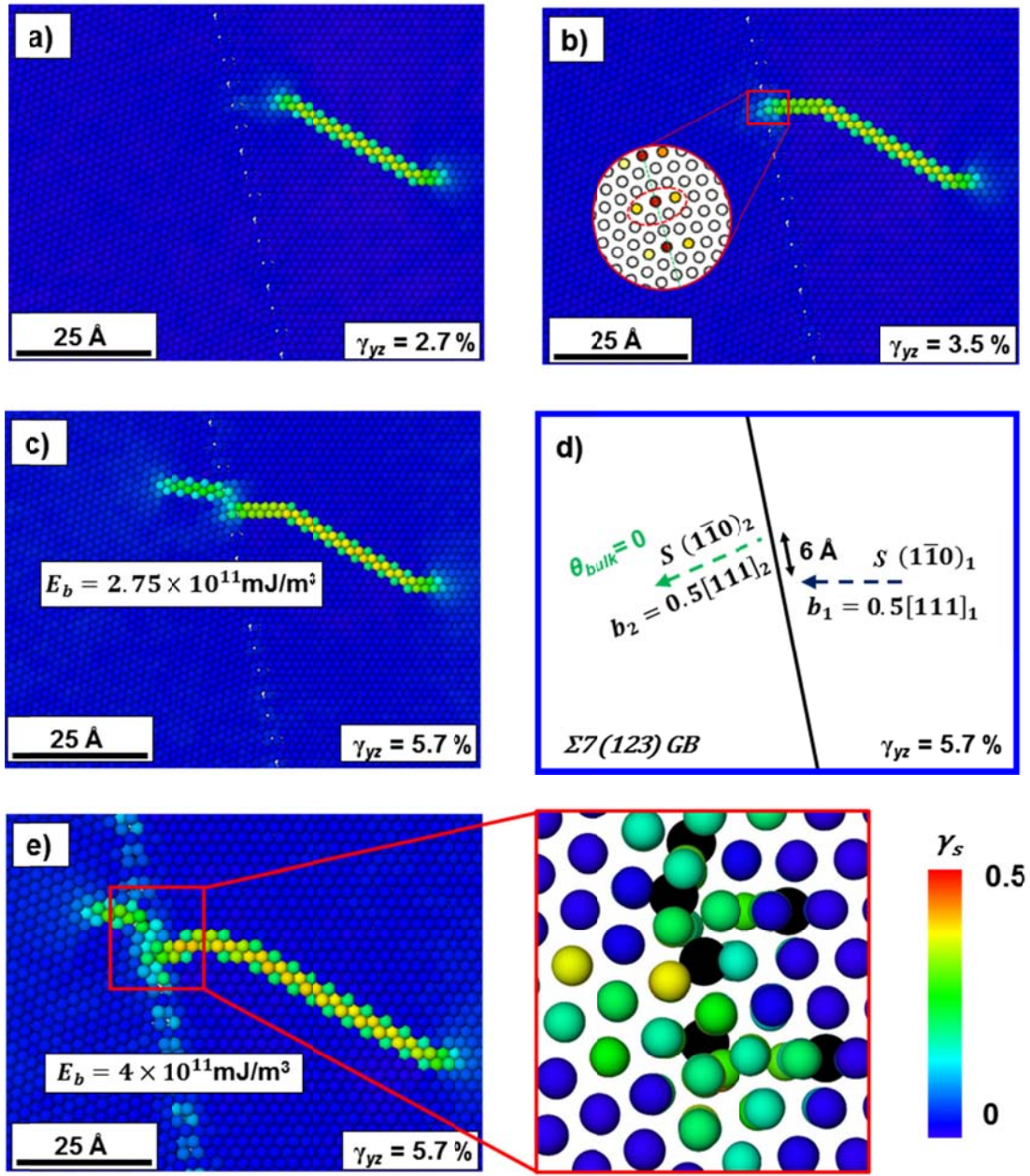

**Figure 3:** a-d) The interaction of a screw dislocation with the Σ7 (123) GB under the applied shear strain ($\gamma_{yz}$) along the Burgers vector direction. The inset b) shows the magnified GB structure colored according to the atomic volume estimated by the Voronoi tessellation as described earlier. d) The schematic representation of the slip transmission mechanism. e) The influence of hydrogen atoms on the DGB interaction. The magnified view in e) shows hydrogen atom positions depicted by black circles. The atoms were colored according to the local atomic shear strain invariant ($\gamma_s$) on a scale of 0 to 0.5.

## 4.2. Σ13 (134) STGB: Change in the transmission mechanism

We now turn to the Σ13 (134) STGB where we observed a change in the transmission mechanism under the influence of a hydrogen rich environment. First, in the case of a hydrogen free environment, the screw dislocation was absorbed in the GB near the site of coincidence (Figure 4a). Subsequently, the dislocation was transmitted along the $(1\bar{1}0)_2$ plane by overcoming an energy barrier of $1.2 \times 10^{11}$ mJ/m$^3$. The predicted outcome using the LRB criterion of the DGB interaction for the Σ13 (134) GB is transmission onto the $(10\bar{1})_2$ plane (refer to Table 1) of the neighboring grain, which is in excellent agreement with the present atomistic work. Further, the dislocation was transmitted from the absorption site at the GB without requiring significant atomic rearrangement of the GB, i.e., a direct transmission.

The slip transmission results from a hydrogen free environment were compared with those obtained using a hydrogen rich environment. The addition of hydrogen along the GB region shows an increase in the energy barrier for slip transmission to $3.5 \times 10^{11}$ mJ/m$^3$ (Figure 4c), i.e., a 192% increase in the energy barrier. The absorbed dislocation was emitted from the GB by an applied shear strain of 8.0% from a site 8 Å away from the initial interaction site (Figure 3c). The presence of hydrogen clearly modifies the DGB interaction (Figure 3c and d) and the outcome, i.e., change in the transmission mechanism (from a direct to indirect transmission).

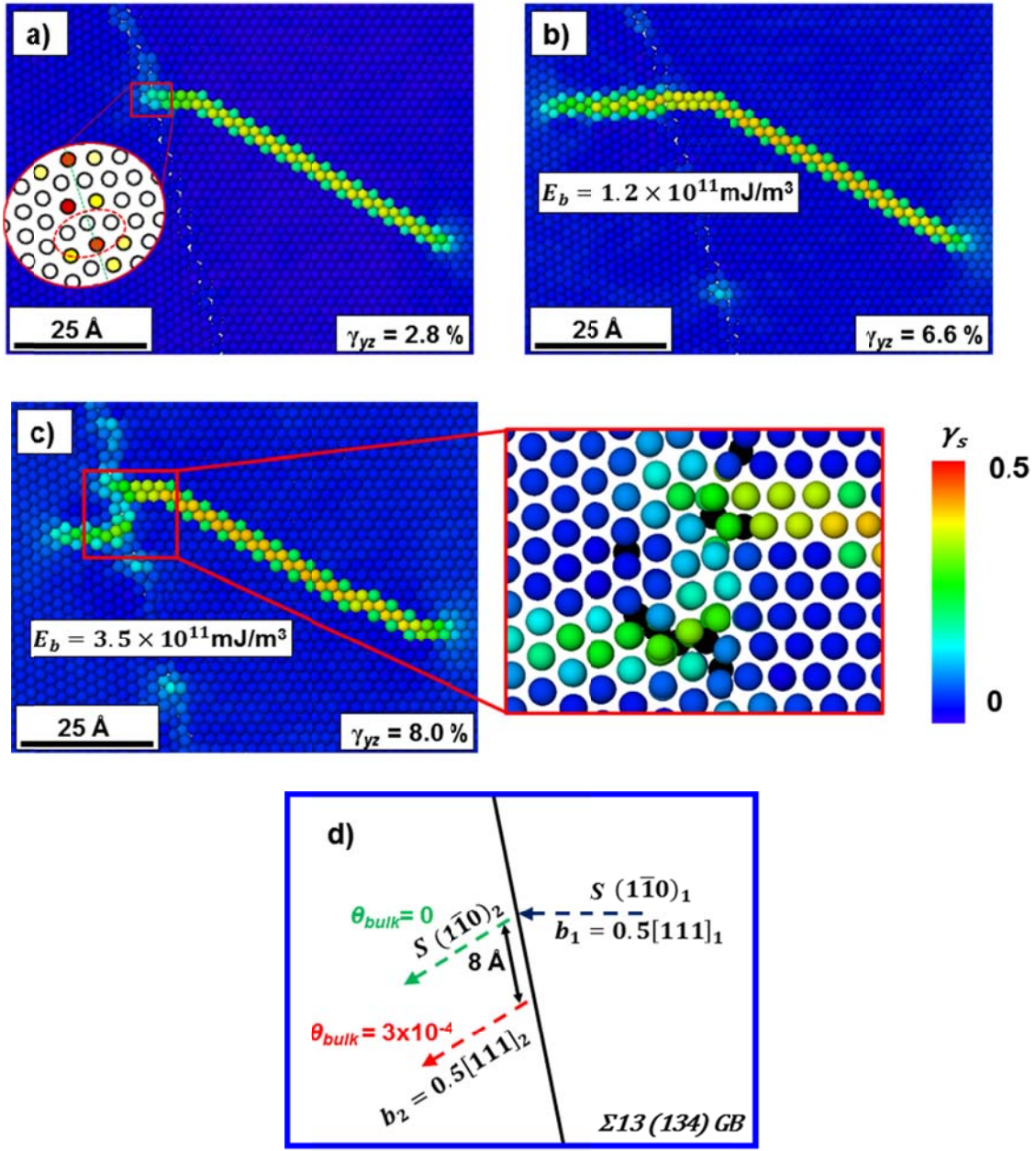

**Figure 4:** a-b) The interaction of a screw dislocation with the Σ13 (134) GB under the applied shear strain ($\gamma_{yz}$) along the Burgers vector direction. The inset a) shows the magnified GB structure colored according to the atomic volume estimated by Voronoi tessellation. c) The change in atomic events for the DGB interaction in the presence of hydrogen. The atoms were colored according to the local atomic shear strain invariant ($\gamma_s$) on a scale of 0 to 0.5. d) The schematic representation of the change in slip transmission mechanism under the hydrogen environment.

### 4.3. Σ57 (178) STGB: Significant change in the distance between the dislocation absorption and transmission

The behavior of slip transfer from the Σ57 GB is an interesting exception to that observed in the previous two GBs, i.e., in the presence of hydrogen we observed very large separation distance between the dislocation absorption and transmission. Initially, in the case of a hydrogen free environment, the screw dislocation was absorbed into the GB. With an increase in the applied shear strain, the dislocation transmits into the neighboring grain by overcoming an energy barrier of $3.3 \times 10^{11}$ mJ/m3 along the $(10\bar{1})_2$ (Figure 5a). This observed behavior is also in good agreement with the analytical prediction (see Table 1) using the LRB criterion. However, with the addition of hydrogen along the GB the slip transmission energy barrier increases to $5.1 \times 10^{11}$ mJ/m3 (Figure 5b). The schematic, as illustrated in Figure 5c, clearly highlights the influence of hydrogen which causes the dislocation to move further away from the absorption site before transmitting into the neighboring grain.

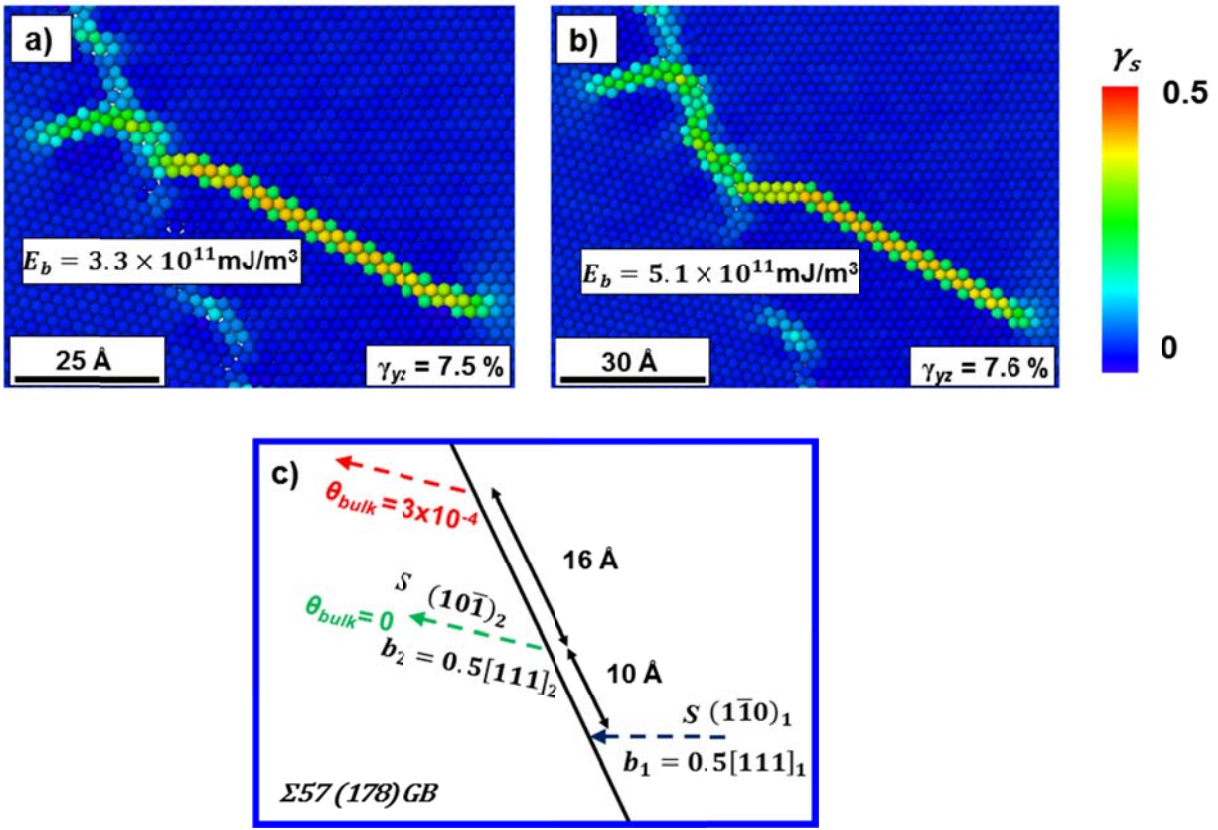

**Figure 5:** a) The interaction of a screw dislocation with the Σ57 (178) GB under the applied shear strain ($\gamma_{yz}$) along the Burgers vector direction. b) The atomic events of hydrogen on the slip transmission mechanism. The atoms were colored according to the local atomic shear strain invariant ($\gamma_s$) on a scale of 0 to 0.5. c) The schematic representation of hydrogen influence on the slip transmission behavior.

### *4.4. Comparison between the predicted and observed slip transmission processes*

For the GBs examined here, it was found that the interface initially pinned the dislocation and the subsequent transmission of the screw dislocation required overcoming the energy barrier by thermal activation or applied stress. Ideally, the dislocation can dissociate into several GB dislocations (GBDs), but this process requires energy for the rearrangement to a GBD network. However, the dislocation dissociation into GBDs with Burger's vector based on the GB primitive vectors [58] reduces the elastic energy at the GB. The outcome of the DGB interaction has been shown to be strongly influenced by the GBD distribution [23,59,60] and the stress state at the interface [22,61]. In the absence of hydrogen, the outcome of the DGB interaction was found to be both direct (Σ13, Σ31, Σ37, and Σ39 GBs) and indirect (Σ7, Σ19, Σ21, and Σ57 GBs) transmission depending on the GBD network. However, the presence of hydrogen atoms along the interface alters the mode of transmission to indirect, except for the Σ13 (134) GB. This can be attributed to modification in the stress field due to the segregated hydrogen atoms along the interface forcing the screw dislocation to travel further away from the initial DGB site in search of a favorable GB site for transmission (Figure 4 and 5).

In general, the outcome of the DGB interaction for both scenarios (with and without hydrogen) was found to be in agreement with the LRB predictions, except for the Σ21 (145), Σ31 (156), and Σ13 (134) GBs where the outgoing slip plane for the dislocation was observed to violate the LRB criterion of maximum resolved shear stress on the slip plane (Table 1). Similar phenomenon has been reported in previous studies [22,23]. The LRB criteria formulated to predict the outcome of the DGB interaction does not encompass either the role of the unique GBD network or the effect of solute atoms/point defects along the interface.

# 5. Effect of hydrogen on dislocation grain boundary interactions: Implications on the fatigue crack initiation

In general, fatigue crack initiation in metals is a direct result of cyclic micro-plasticity, specifically cyclic irreversibility associated with the glide of dislocations. A critical feature of cyclic plastic deformation is the formation of complex localized surface reliefs (persistent slip markings) at the traces of emerging persistent slip bands (PSBs). GBs offer resistance to the easy glide of dislocations, resulting in dislocation pile-ups. The strain localization at the dislocation pile-up is relieved by the transfer of dislocations across the GB. However, the presence of an impenetrable GB can generate localization of strain that could eventually lead to a transgranular crack initiation [3], i.e., formation of slip bands during fatigue loading conditions that eventually become shear cracks. Further, under an aggressive environment, such as the case in hydrogen embrittlement, the segregation of adequate hydrogen atoms along the GBs leads to a reduction in the cohesive strength of the interface [6,7]. These condition increases the susceptibility for an intergranular failure at the dislocation pile-up [15]. Thus, the computed energy barrier for slip transmission is discussed within the continuum energy balance description for a crack initiation due to dislocation pile-ups. See the work of Tanaka et al. [30] for further details.

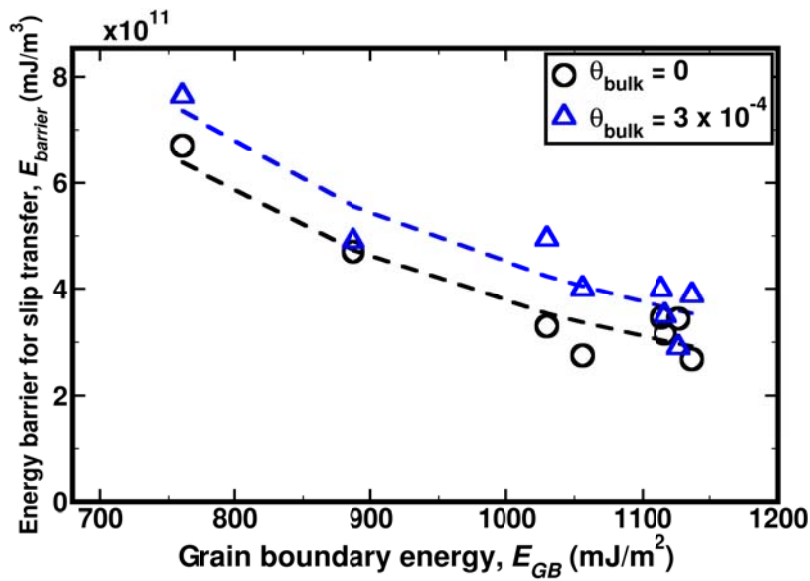

**Figure 6:** The effect of hydrogen on the slip transmission barrier through the <111> STGBs.

The energy balance approach takes into consideration the energy contributions of a) the PSB structure under the applied stress ($E_{PSB}$) [30,31]; and b) the interaction energy of the PSB with the GB ($E_{pen}$) which represents the fraction of the dislocation density penetrating/transmitting across the GB. For the crack to instigate, the stored strain energy at the dislocation pileup site ($E_{int}$) must be greater than the energy required to nucleate an intergranular crack at the GB ($E_{cleavage}$) which can be formulated in the following manner:

$$E_{int} = E_{PSB}(\sigma_a, N, \rho, \tau_c, \theta) + E_{pen}(N, \rho, E_{barrier}, \theta) \tag{5a}$$

$$E_{int} \geq E_{cleavage}(\gamma_s^{GB}, \theta) \tag{5b}$$

where $\sigma_a$ is the applied stress, $N$ is the number of fatigue cycles, $\rho$ is the dislocation density within the PSB structure, $\tau_c$ is the Peierls stress, $\theta$ is the hydrogen concentration at the GB calculated using the atomic occupancy of hydrogen (Equation 3), $E_{barrier}$ is the energy barrier for slip transmission across the GB, and $E_{GB}$ is the GB energy. The presence of hydrogen increases the glissile dislocation density ($\rho_{0,H}$) in comparison with a hydrogen free ($\rho_0$) condition because of the shielding effect induced by the hydrogen environment surrounding a dislocation [27,28,62]. The interaction energy contribution of the PSB with the GB ($E_{pen}$) is directly related to the energy barrier for a single dislocation to transmit across the GB interface ($E_{barrier}$). Using atomistic simulations, a strong inverse relationship between the energy barrier for slip transmission ($E_{barrier}$) and the initial GB energy ($E_{GB}$) was found (Figure 6). In other words, GBs with lower interfacial energy demonstrated a higher barrier for slip transmission, which is in agreement with previous findings on Ni [24]. An inverse power law fit was found to best describe this relation.

$$E_{barrier} = \alpha(E_{GB})^\beta \tag{6}$$

where $\alpha$ and $\beta$ are model parameters, which are listed in Table 2. The presence of hydrogen along the GB consistently increases the energy barrier for slip transmission across the <111> STGBs (Figure 6). Further, at room temperature, the hydrogen atoms diffuse along with the glissile dislocations [27], causing a redistribution of hydrogen concentration towards sites of strain localization, i.e., the dislocation-GB pile-up site acts as sink for hydrogen atoms. On the other hand, it has been well established in literature [6,7] that increasing hydrogen concentration along the GB reduces the GB cohesive strength (hydrogen enhanced decohesion mechanism).

**Table 2:** Model parameters related to an inverse power law fit between the energy barrier for slip transmission and the initial GB energy.

| Bulk hydrogen concentration | $\alpha$ | $\beta$ |
| --- | --- | --- |
| $\theta_{bulk} = 0$ | $2.54 \times 10^{17}$ | $-1.94$ |
| $\theta_{bulk} = 3 \times 10^{-4}$ | $1.29 \times 10^{17}$ | $-1.82$ |

Thus, based on previous experimental studies [14,51,53] and this work, a case for hydrogen induced plasticity mediated intergranular fatigue crack initiation is presented. In other words, the presence of hydrogen triggers a series of microstructural events that increase the localized strain energy stored within the GB ($E_{int}$). Further, providing adequate energy to activate alternative relief mechanisms in the absence of slip transmission, such as hydrogen induced intergranular decohesion. Hence, based on atomistic findings and the examination of energy balance for a fatigue crack initiation, we highlight the role of plasticity in hydrogen-induced intergranular failure (for α-Fe). This is in agreement with previous experimental studies [14,64,65] that have suggested similar significance of plasticity on the hydrogen embrittlement mechanism. The effect of hydrogen on the increased susceptibility of the microstructure to intergranular fatigue crack initiation can be summarized in terms of the schematic in Figure 7.

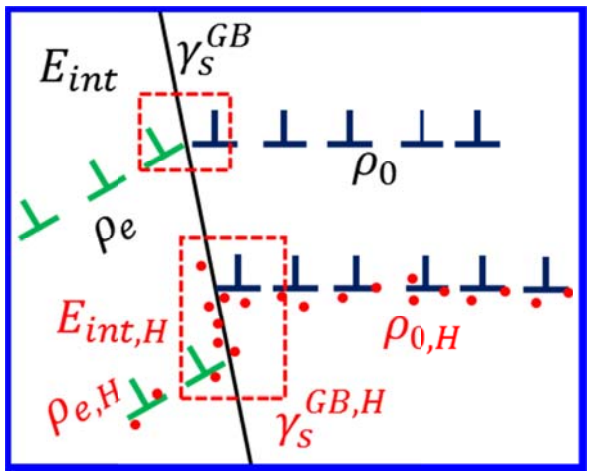

**Figure 7:** Schematic representation of the effect of hydrogen on the intergranular fatigue crack initiation. The top half shows the hydrogen free interaction of incident dislocation density of $\rho_0$ and the emitted dislocation density is $\rho_e$. In the presence of hydrogen, the incident dislocation density $(\rho_{0,H})$ increases in comparison to the hydrogen free dislocation density $(\rho_0)$ because of the shielding effect afforded by the hydrogen environment surrounding a dislocation. The emitted dislocation density $(\rho_{e,H})$ is lower in than the hydrogen free case $(\rho_{e,H} < \rho_e)$

## 6. Conclusions

In this work, the role of hydrogen on the DGB interaction was examined for several <111> STGBs in α-Fe using atomistic simulations. The primary objectives of this work were: a) to understand the role of atomic structure at the DGB interaction site on slip transmission mechanisms, b) quantify the influence of hydrogen on the DGB interactions, and c) reexamine the continuum fatigue crack initiation criterion in light of these findings. The significant contributions of this study are as follows:

1. The outcome of the DGB interaction was found to strongly depend on the GB character for the cases with or without the hydrogen rich environment at the interface. This was illustrated by comparing the observed outcome of the DGB interaction with the LRB criteria predictions. The LRB criteria predicts the outcome of DGB interaction by relying entirely on the slip system geometry and the stress state. Thus, in some cases such as the Σ21 (145), Σ31 (156), and Σ13 (134) GBs, the LRB criteria was found to be unable to predict the outgoing slip plane. Hence, further modifications to the LRB criteria are needed such as the effect of GB dislocation network on the slip transmission mechanism.

2. In several interfaces examined here, the presence of a hydrogen rich environment along the interface modifies the DGB interaction by forcing the dislocation to move further away from the absorption site before transmitting into the neighboring grain. In the case of the Σ13 (134) GB, the segregated hydrogen atoms alter the slip transmission mechanism, i.e. from a direct to an indirect manner.

3. The segregated hydrogen atoms along the GB distort the underlying structure, thereby consistently increasing the energy barrier for slip transmission. This leads to an increase in the dislocation density within the dislocation pile-up.

4. The energy required for the dislocation to transmit across the GB was found to be inversely related to the initial GB energy in both environments, i.e., hydrogen-free and hydrogen charged. Thus it was found that the GBs with lower interfacial energy demonstrated a higher barrier for slip transmission.

5. Finally, based on the present work and previous experimental studies [6,7,14], the energy balance for a fatigue crack initiation was examined by considering the various interaction energies under the influence of hydrogen. The presence of hydrogen was found to increase the energy barrier required for dislocations trapped at the interface to transmit across the interface. In addition, the plasticity provides an effective transport medium for hydrogen to be deposited at the GB [14] (HELP) thereby reducing the GB cohesive strength [6,7]. Overall, the absence of a feasible reliving mechanism for the trapped dislocations establishes conditions for a transition in the failure mechanism, i.e., a transition from transgranular-to-intergranular fracture mode. Thus, this suggests a plasticity mediated intergranular failure as possible outcome for fatigue failure in a hydrogen rich environment.

In summary, this systematic study quantifies the increased energy barrier for slip transmission due to the presence of hydrogen at the interface. Further, evidence was presented supporting the notion that during hydrogen embrittlement, failure occurs in an intergranular manner, but the circumstances for this are created by the HELP mechanism and the inability of slip transmission across the GB. However, we note that the dislocation–hydrogen interaction is strongly dependent on the dislocation pile-up configurations [28]. Therefore, investigating the barrier for slip transmission for various dislocation densities is needed in the future. Nonetheless, our study provides critical knowledge towards a comprehensive understanding of hydrogen effects on the slip transmission (in general deformation) behaviors of $\alpha$-Fe.

**Acknowledgement**

The authors gratefully acknowledge support from the Office of Naval Research under contract N00014111079.